\documentclass[aps,prl,twocolumn,superscriptaddress,showpacs,amsmath,amssymb]{revtex4}
\usepackage{graphicx}
\usepackage{dcolumn} 
\usepackage{bm}      
\usepackage{ifpdf}   
\usepackage{color}   
\usepackage{dcolumn} 

\ifpdf
   \pdfoutput=1           
   \pdftrue               
\else
   \pdfoutput=0
   \pdffalse              
\fi

\ifpdf
  \DeclareGraphicsExtensions{.pdf,.jpg,.jpeg}
  \definecolor{rltred}{rgb}{0.75,0,0}
  \definecolor{rltgreen}{rgb}{0,0.5,0}
  \definecolor{rltblue}{rgb}{0,0,0.75}
  \usepackage[colorlinks%
  ,bookmarks%
  ,urlcolor=rltblue%
  ,filecolor=rltgreen%
  ,linkcolor=rltred%
  ,pdftitle={W polarization measurement in dilepton channel at CDF}%
  ,pdfsubject={W polarization measurement in dilepton channel at CDF with
    5.1 fb}%
  ,pdfkeywords={top quark, dilepton channel, W polarization  measurement}%
  ,pdfauthor={Roman Lysak <lysak@saske.sk>}%
  ,pdfpagemode={UseOutlines}%
  ,bookmarksopen=true%
  ,bookmarksopenlevel=1%
  ,bookmarksnumbered=true%
  ,pdfstartview={Fit}%
  ]{hyperref}
\else
  \DeclareGraphicsExtensions{.eps,.ps,.eps.gz,.ps.gz}
\fi
 
\newcommand{\MET}{/\!\!\!\!E_T}
\newcommand{\ct}{$\cos \theta^*\ $}
\def\GeVc2{\ensuremath{\mathrm{ Ge\kern -0.1em V }\kern -0.2em /c^2 }}

\begin{document}


\preprint{Draft 3.1}


\title{W-boson polarization measurement in the $t\bar{t}$ dilepton channel
  using the CDF II Detector} 

\affiliation{Institute of Physics, Academia Sinica, Taipei, Taiwan 11529, Republic of China}
\affiliation{Argonne National Laboratory, Argonne, Illinois 60439, USA}
\affiliation{University of Athens, 157 71 Athens, Greece}
\affiliation{Institut de Fisica d'Altes Energies, ICREA, Universitat Autonoma de Barcelona, E-08193, Bellaterra (Barcelona), Spain}
\affiliation{Baylor University, Waco, Texas 76798, USA}
\affiliation{Istituto Nazionale di Fisica Nucleare Bologna, $^{ee}$University of Bologna, I-40127 Bologna, Italy}
\affiliation{University of California, Davis, Davis, California 95616, USA}
\affiliation{University of California, Los Angeles, Los Angeles, California 90024, USA}
\affiliation{Instituto de Fisica de Cantabria, CSIC-University of Cantabria, 39005 Santander, Spain}
\affiliation{Carnegie Mellon University, Pittsburgh, Pennsylvania 15213, USA}
\affiliation{Enrico Fermi Institute, University of Chicago, Chicago, Illinois 60637, USA}
\affiliation{Comenius University, 842 48 Bratislava, Slovakia; Institute of Experimental Physics, 040 01 Kosice, Slovakia}
\affiliation{Joint Institute for Nuclear Research, RU-141980 Dubna, Russia}
\affiliation{Duke University, Durham, North Carolina 27708, USA}
\affiliation{Fermi National Accelerator Laboratory, Batavia, Illinois 60510, USA}
\affiliation{University of Florida, Gainesville, Florida 32611, USA}
\affiliation{Laboratori Nazionali di Frascati, Istituto Nazionale di Fisica Nucleare, I-00044 Frascati, Italy}
\affiliation{University of Geneva, CH-1211 Geneva 4, Switzerland}
\affiliation{Glasgow University, Glasgow G12 8QQ, United Kingdom}
\affiliation{Harvard University, Cambridge, Massachusetts 02138, USA}
\affiliation{Division of High Energy Physics, Department of Physics, University of Helsinki and Helsinki Institute of Physics, FIN-00014, Helsinki, Finland}
\affiliation{University of Illinois, Urbana, Illinois 61801, USA}
\affiliation{The Johns Hopkins University, Baltimore, Maryland 21218, USA}
\affiliation{Institut f\"{u}r Experimentelle Kernphysik, Karlsruhe Institute of Technology, D-76131 Karlsruhe, Germany}
\affiliation{Center for High Energy Physics: Kyungpook National University, Daegu 702-701, Korea; Seoul National University, Seoul 151-742, Korea; Sungkyunkwan University, Suwon 440-746, Korea; Korea Institute of Science and Technology Information, Daejeon 305-806, Korea; Chonnam National University, Gwangju 500-757, Korea; Chonbuk National University, Jeonju 561-756, Korea}
\affiliation{Ernest Orlando Lawrence Berkeley National Laboratory, Berkeley, California 94720, USA}
\affiliation{University of Liverpool, Liverpool L69 7ZE, United Kingdom}
\affiliation{University College London, London WC1E 6BT, United Kingdom}
\affiliation{Centro de Investigaciones Energeticas Medioambientales y Tecnologicas, E-28040 Madrid, Spain}
\affiliation{Massachusetts Institute of Technology, Cambridge, Massachusetts 02139, USA}
\affiliation{Institute of Particle Physics: McGill University, Montr\'{e}al, Qu\'{e}bec, Canada H3A~2T8; Simon Fraser University, Burnaby, British Columbia, Canada V5A~1S6; University of Toronto, Toronto, Ontario, Canada M5S~1A7; and TRIUMF, Vancouver, British Columbia, Canada V6T~2A3}
\affiliation{University of Michigan, Ann Arbor, Michigan 48109, USA}
\affiliation{Michigan State University, East Lansing, Michigan 48824, USA}
\affiliation{Institution for Theoretical and Experimental Physics, ITEP, Moscow 117259, Russia}
\affiliation{University of New Mexico, Albuquerque, New Mexico 87131, USA}
\affiliation{The Ohio State University, Columbus, Ohio 43210, USA}
\affiliation{Okayama University, Okayama 700-8530, Japan}
\affiliation{Osaka City University, Osaka 588, Japan}
\affiliation{University of Oxford, Oxford OX1 3RH, United Kingdom}
\affiliation{Istituto Nazionale di Fisica Nucleare, Sezione di Padova-Trento, $^{ff}$University of Padova, I-35131 Padova, Italy}
\affiliation{University of Pennsylvania, Philadelphia, Pennsylvania 19104, USA}
\affiliation{Istituto Nazionale di Fisica Nucleare Pisa, $^{gg}$University of Pisa, $^{hh}$University of Siena and $^{ii}$Scuola Normale Superiore, I-56127 Pisa, Italy}
\affiliation{University of Pittsburgh, Pittsburgh, Pennsylvania 15260, USA}
\affiliation{Purdue University, West Lafayette, Indiana 47907, USA}
\affiliation{University of Rochester, Rochester, New York 14627, USA}
\affiliation{The Rockefeller University, New York, New York 10065, USA}
\affiliation{Istituto Nazionale di Fisica Nucleare, Sezione di Roma 1, $^{jj}$Sapienza Universit\`{a} di Roma, I-00185 Roma, Italy}
\affiliation{Rutgers University, Piscataway, New Jersey 08855, USA}
\affiliation{Texas A\&M University, College Station, Texas 77843, USA}
\affiliation{Istituto Nazionale di Fisica Nucleare Trieste/Udine, I-34100 Trieste, $^{kk}$University of Udine, I-33100 Udine, Italy}
\affiliation{University of Tsukuba, Tsukuba, Ibaraki 305, Japan}
\affiliation{Tufts University, Medford, Massachusetts 02155, USA}
\affiliation{University of Virginia, Charlottesville, Virginia 22906, USA}
\affiliation{Waseda University, Tokyo 169, Japan}
\affiliation{Wayne State University, Detroit, Michigan 48201, USA}
\affiliation{University of Wisconsin, Madison, Wisconsin 53706, USA}
\affiliation{Yale University, New Haven, Connecticut 06520, USA}

\author{T.~Aaltonen}
\affiliation{Division of High Energy Physics, Department of Physics, University of Helsinki and Helsinki Institute of Physics, FIN-00014, Helsinki, Finland}
\author{B.~\'{A}lvarez~Gonz\'{a}lez$^z$}
\affiliation{Instituto de Fisica de Cantabria, CSIC-University of Cantabria, 39005 Santander, Spain}
\author{S.~Amerio}
\affiliation{Istituto Nazionale di Fisica Nucleare, Sezione di Padova-Trento, $^{ff}$University of Padova, I-35131 Padova, Italy}
\author{D.~Amidei}
\affiliation{University of Michigan, Ann Arbor, Michigan 48109, USA}
\author{A.~Anastassov$^x$}
\affiliation{Fermi National Accelerator Laboratory, Batavia, Illinois 60510, USA}
\author{A.~Annovi}
\affiliation{Laboratori Nazionali di Frascati, Istituto Nazionale di Fisica Nucleare, I-00044 Frascati, Italy}
\author{J.~Antos}
\affiliation{Comenius University, 842 48 Bratislava, Slovakia; Institute of Experimental Physics, 040 01 Kosice, Slovakia}
\author{G.~Apollinari}
\affiliation{Fermi National Accelerator Laboratory, Batavia, Illinois 60510, USA}
\author{J.A.~Appel}
\affiliation{Fermi National Accelerator Laboratory, Batavia, Illinois 60510, USA}
\author{T.~Arisawa}
\affiliation{Waseda University, Tokyo 169, Japan}
% ...

\collaboration{CDF Collaboration\footnote{With visitors from
$^a$Istituto Nazionale di Fisica Nucleare, Sezione di Cagliari, 09042 Monserrato (Cagliari), Italy,
$^b$University of CA Irvine, Irvine, CA 92697, USA,
$^c$University of CA Santa Barbara, Santa Barbara, CA 93106, USA,
$^d$University of CA Santa Cruz, Santa Cruz, CA 95064, USA,
$^e$Institute of Physics, Academy of Sciences of the Czech Republic, Czech Republic,
$^f$CERN, CH-1211 Geneva, Switzerland,
$^g$Cornell University, Ithaca, NY 14853, USA,
$^h$University of Cyprus, Nicosia CY-1678, Cyprus,
$^i$Office of Science, U.S. Department of Energy, Washington, DC 20585, USA,
$^j$University College Dublin, Dublin 4, Ireland,
$^k$ETH, 8092 Zurich, Switzerland,
$^l$University of Fukui, Fukui City, Fukui Prefecture, Japan 910-0017,
$^m$Universidad Iberoamericana, Mexico D.F., Mexico,
$^n$University of Iowa, Iowa City, IA 52242, USA,
$^o$Kinki University, Higashi-Osaka City, Japan 577-8502,
$^p$Kansas State University, Manhattan, KS 66506, USA,
$^q$Korea University, Seoul, 136-713, Korea,
$^r$University of Manchester, Manchester M13 9PL, United Kingdom,
$^s$Queen Mary, University of London, London, E1 4NS, United Kingdom,
$^t$University of Melbourne, Victoria 3010, Australia,
$^u$Muons, Inc., Batavia, IL 60510, USA,
$^v$Nagasaki Institute of Applied Science, Nagasaki, Japan,
$^w$National Research Nuclear University, Moscow, Russia,
$^x$Northwestern University, Evanston, IL 60208, USA,
$^y$University of Notre Dame, Notre Dame, IN 46556, USA,
$^z$Universidad de Oviedo, E-33007 Oviedo, Spain,
$^{aa}$CNRS-IN2P3, Paris, F-75205 France,
$^{bb}$Texas Tech University, Lubbock, TX 79609, USA,
$^{cc}$Universidad Tecnica Federico Santa Maria, 110v Valparaiso, Chile,
$^{dd}$Yarmouk University, Irbid 211-63, Jordan,
}}
\noaffiliation

\date{\today}

\begin{abstract}
We present a measurement of the $W$-boson polarization in top-quark decays
in $t\bar{t}$ events with decays to dilepton final states using data corresponding 
to $5.1~{\rm fb^{-1}}$ of integrated luminosity in $p\bar{p}$ collisions
collected by the CDF II detector at the Tevatron.  
Assuming a top-quark mass of 172.5 GeV/c$^2$, a simultaneous measurement of the 
fractions of longitudinal ($f_0$) and
right-handed ($f_+$) $W$-bosons yields the results $f_0 = 0.70
^{+0.18}_{-0.17} {\rm  (stat)} \pm 0.06 {\rm (syst)}$ and $f_+ = -0.09
\pm 0.09 {\rm  (stat)} \pm 0.03 {\rm (syst)}$. 
Combining this measurement with our previous measurement based on single-lepton
 final states, we obtain  $f_0 =  0.84 \pm 0.09 {\rm (stat)} \pm
0.05 {\rm (syst)}$ and $f_{+} =  -0.16 \pm 0.05 {\rm (stat)} \pm 0.04
{\rm (syst)}$. 
The results are consistent with the standard model expectation. 
\end{abstract}


\author{The CDF collaboration}

\maketitle

%
%
Since the top quark discovery by the CDF and D0 experiments in 1995,
many of its properties have been measured. Due to its very short
lifetime~\cite{lifetimeCDF, lifetimeD0}, the top quark does not hadronize and therefore its
properties are transferred directly to its decay products.   
The standard model (SM) makes specific predictions for the
$W$-boson polarization in top-quark decays. Precise measurement of the
$W$-boson polarization provides a test of the SM and
could reveal new physics beyond the SM~\cite{WhelBeyondSM}.  

In the SM, the top quark decays to a $W$-boson and $b$ quark with
almost 100\% probability~\cite{pdg}. The $W$-boson is a massive vector
particle with three polarization states: 
right-handed (+1), longitudinal (0), and left-handed \mbox{(--1)}. 
The di\-ffe\-rential decay rate of the top quark is given by:
\begin{equation}
\footnotesize
  \frac{d\Gamma}{d \cos \theta^{*}}  \propto  
   (1 -  \cos \theta^{*})^{2} f_{-} + 2 (1 - {\cos}^2 \theta^{*}) f_{0} + (1 + \cos \theta^{*})^{2} f_{+}
\label{eq:cosThetaTheory}
\end{equation}  
where $\theta^{*}$ is the angle between the direction 
of the charged lepton (or down-type quark from the decay of the $W$-boson) and the opposite
direction of the top quark in the $W$-boson rest frame. The polarization
fractions satisfy the normalization condition $f_- + f_0 + f_+ = 1$.
The right-handed $W$-boson production in the context of the SM is strongly
suppressed due to the $V-A$ structure of the charged-current weak interaction.
In the SM at the tree level~\cite{WhelBeyondSM},
the fraction of right-handed  $W$-bosons is very close to zero ($f_+ =
3.7 \times 10^{-4}$) while $f_{0}=0.698$ and $f_{-}=0.301$ for a
top-quark mass ($m_{top}$) of 173.3 \GeVc2~\cite{cdfd0_topMassCombination},
a $W$-boson mass ($m_W$) of 80.4 \GeVc2~\cite{pdg}, and a $b$-quark mass ($m_b$) of 4.78
\GeVc2~\cite{pdg}. 
In the $m_b \rightarrow 0$ limit,  $f_+ = 0$ and $f_{0} =  m_{{\rm top}}^{2}
/ \left( 2m_{W}^{2}+m_{{\rm top}}^{2} \right)$. 
Polarization fractions that deviate from the SM values are predicted in
theories with anomalous $tWb$ couplings~\cite{WhelBeyondSM}.

An earlier measurement of the polarization fractions of the $W$-boson in
top-quark decays by the CDF collaboration focused on the 
single-lepton channel~\cite{whelicityCDF}. 
Results have also been reported by the D0 and ATLAS collaborations, 
where the measurements in the single-lepton and dilepton channels are 
combined~\cite{whelicityD0, whelicityATLAS}
In this Letter we
report the CDF measurement in the dilepton channel ($t\bar{t} \rightarrow
W^{+}b W^{-}\bar{b} \rightarrow \ell^{+}\nu b \ell^{-}\bar{\nu}
\bar{b}$) where $\ell=e,\mu$ and leptonically decaying $\tau$ leptons
are also included into the signal.
We perform two types of measurements: a model-independent approach
where $f_0$ and $f_{+}$ are determined simultaneously; and a
model-dependent approach where $f_0$ ($f_{+}$) is fixed to its SM
value, and $f_{+}$ ($f_0$) is measured. 
The model-independent and model-dependent approaches are referred as
``2D" and ``1D", respectively, throughout this article.   
We also combine this result with our previous
measurement~\cite{whelicityCDF} in the single-lepton channel.   

%
%

This analysis is based on data corresponding to an integrated luminosity
of $5.1\ {\rm    fb^{-1}}$ collected with the CDF II
detector~\cite{CdfDetector} between March 2002 and June 2009 at the
Fermilab Tevatron with a center of mass energy $\sqrt{s}$ = 1.96 TeV.   
The CDF II detector is described in detail elsewhere~\cite{CdfDetector}.
The components essential to this analysis are the tracking system
consisting of a silicon microstrip tracker and a central drift chamber
immersed in a 1.4 T solenoidal magnetic field, electromagnetic and
hadronic calorimeters arranged in a projective geometry outside the
magnet coil, and drift chambers and scintillation counters for muon
detection outside the calorimeters.

The analysis uses the same event selection criteria that were used
for the measurement of the $t\bar{t}$ cross-section in the dilepton
channel~\cite{CDFtop_xs_dil}. A brief description of the event
selection is as follows: 
the data were collected with an online inclusive event-selection
system (trigger) that requires 
a lepton (electron or muon) with transverse energy $E_T>18$ GeV.
From the inclusive lepton data, we select events with oppositely-charged leptons 
%
%
%
 of $E_T > 20\ {\rm GeV}$. 
We define leptons to satisfy very restrictive identification and isolation criteria as ``tight''
while ``loose'' leptons satisfy less restrictive identification criteria and has no isolation requirement.
We require at least one tight lepton in each dilepton candidate.
 The pseudorapidity $\eta$~\cite{transverseEt} coverage is $|\eta| <
 2.0$ for electrons and $|\eta| < 1.0$ for muons. 
 We require missing transverse energy $\MET > 25\ {\rm
  GeV}$~\cite{missingTransverseEt} unless the $\MET$ direction is along
(within $20^{\circ}$ in $\phi$) either a lepton or a jet, in which
case we require $\MET > 50\ {\rm GeV}$. 
Additionally we require at least two jets~\cite{jetDefinition}
reconstructed with $E_T > 15\ {\rm GeV}$ and $|\eta| < 2.5$. Jet
energies are corrected for the effects of calorimeter  
response, multiple interactions, and the hadronic calorimeter energy 
scale~\cite{JES_NIM}.   
Background contributions are further reduced through kinematic cuts 
on the dilepton invariant mass, total energy in the transverse plane 
and $\MET$ significance~\cite{CDFtop_xs_dil}. 
In this measurement, we split the inclusive data sample into two
non-overlapping subsamples (``$b$-tag'' and ``non-tag'') where for the 
``$b$-tag'' subsample we require at least one jet in the event 
to be consistent with having originated from a $b$ quark by using an
algorithm that identifies a long-lived $b$ hadron through the presence of a
displaced vertex~\cite{cdf_btagging}. The two subsamples  have 
different signal-to-background ratios and background compositions;
therefore, we can improve the overall measurement uncertainties by
analyzing each subsample separately.

The dominant background contribution is from ``fake''
events where a jet is misidentified as a lepton. The main source of
``fake'' events is $W (\rightarrow \ell \nu)$ + jets events.
An additional background is the Drell-Yan production of electrons or 
muons ($q\bar{q} \rightarrow Z/\gamma^{*} \rightarrow
\ell^{+}\ell^{-}$, where $\ell =e,\mu$) with spurious $\MET$. Both of the
above background contributions are estimated using data-based methods~\cite{CDFtop_xs_dil}.
The remaining background contributions are from the Drell-Yan production of $\tau$
leptons  and diboson ($WW$,$WZ$,$ZZ$) production which are
estimated using Monte-Carlo (MC) simulation.  
 The detailed description of the background estimation can be found
 in~\cite{CDFtop_xs_dil}.  
The numbers of expected and observed events passing dilepton selection
are given in Table~\ref{tab:final_bkgr_5.1fb}. Statistical and systematic
uncertainties are combined for the numbers of events. 
The correlations among signal and backgrounds systematic uncertainties are 
considered for the total numbers of events.
There is a good agreement between data and expectations from $t\bar{t}$ production and
background contributions (this holds also for ``$b$-tag''
and ``non-tag'' subsamples individually).  

\begin{table}
  \caption{  Expected and observed numbers of signal and background
    events after the dilepton selection assuming $\sigma_{t\bar{t}} =
    6.7$ pb ($m_{top} = 175$ \GeVc2).}  
  \label{tab:final_bkgr_5.1fb}
  \begin{center}
    \begin{ruledtabular}
    \begin{tabular}{lccc}
      \multicolumn{1}{c}{ Source  }             &  \multicolumn{3}{c}{ Events  }   \\
      \hline 
      Diboson                                             &  17.5 & $\pm$ & 4.6 \\      
      $Z/\gamma^* \rightarrow \tau\tau$                   &  12.3 & $\pm$ & 2.2 \\
      $Z/\gamma^* \rightarrow ee +\mu\mu$                 &  22.4 & $\pm$ & 3.2 \\
      Fakes                                               &  53.7 & $\pm$ & 14.7\\
      \hline                                                      
      \multicolumn{1}{l}{Total background}                & 105.8 & $\pm$ & 17.2\\
      \multicolumn{1}{l}{$t\bar{t}$ ($\sigma = 6.7$ pb)}  & 222.4 & $\pm$ & 10.6\\
      \hline                                              
      Total SM expectation                                & 328.2 & $\pm$ & 27.6\\
      \hline                                        
      \multicolumn{1}{l}{ Observed  }                     & \multicolumn{3}{c}{ 343 }\\
    \end{tabular}
    \end{ruledtabular}
  \end{center}
\end{table}

%
%
We use the \ct of the leptons defined above to determine the $W$-boson
polarization 
fractions. In order to reconstruct $\cos \theta^*$, the full
event kinematics of $t\bar{t}$ decay chain must be reconstructed. The dilepton 
channel presents an under-constrained system due to the two undetected
neutrinos. We use a simple modification of the kinematic method
previously developed for the measurement of the top-quark mass, denoted as
the KIN method in~\cite{weTopMassPRD}. We solve the kinematic
equations by Newton's method for nonlinear systems of equations using
the top-quark mass constraint with $m_{top} = 175$ \GeVc2.  
In principle each event provides two measurements of  $\cos \theta^*$.
However, some events do not have kinematic solutions under our assumptions, 
and these events are discarded.

\begin{figure}[htbp]
  \begin{center}
    \includegraphics[width=3.375in]{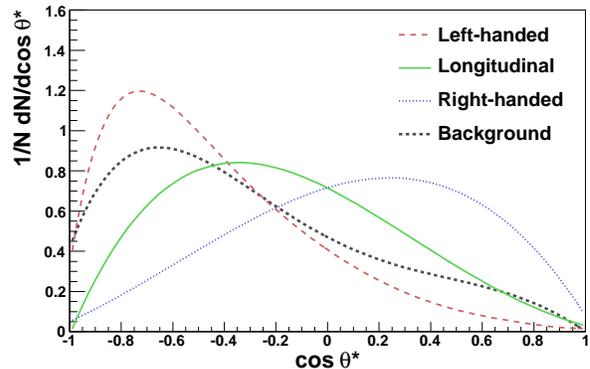}
  \end{center} 
  \caption{The signal templates for left-handed, longitudinal and
    right-handed $W$-bosons together with the background template for
    the inclusive dilepton selection.}  
  \label{fig:templates}
\end{figure}

In order to perform a fit to the $\cos \theta^*$ distribution we
create templates using $t\bar{t}$ MC simulated samples
for exclusive  
left-handed, longitudinal and right-handed $W$-bosons using a
customized \textsc{herwig}~\cite{ggwig, herwig} MC generator.  
We create the templates separately for ``$b$-tag'' and ``non-tag''
subsamples which turn out to be similar. 
Figure~\ref{fig:templates} shows the templates for the inclusive
sample for both the signal and background.  

Due to various selection and reconstruction effects 
(e.g., we consider
the two highest $E_T$ jets as jets coming from $b$-quark hadronization
while there is a possibility that one of these two jets comes from
 initial (ISR) or final (FSR) state gluon radiation) the
templates vary significantly from the theoretical distributions in 
Eq.~(\ref{eq:cosThetaTheory}).  The effect of the lepton $E_T$ and
the isolation cut is 
seen as a softening of the theoretical peaks near $\cos \theta^*
= -1$.  Furthermore, the KIN reconstruction method requires the 
lepton-jet-$\MET$ mass to be close to the mass of the top quark so that the
reconstruction is inefficient for high lepton-jet pair masses ($\cos
\theta^* \simeq +1$). This gives a polarization-dependent
reconstruction efficiency of about 95\%, 92\%, and 87\%, respectively, for
left-handed, longitudinal, and  right-handed $W$-bosons.
For the background events, the reconstruction efficiency is only 71\%.
There are also differences in the acceptance of dilepton events.
The difference can be as large as about 30\%, as observed between events 
with two left-handed W bosons and those with two longitudinal W bosons, 
where the latter has a relatively larger acceptance than the former.
This is mainly due  to the dependence of the acceptance on the lepton 
$p_T$ and isolation (leptons from left-handed $W$-bosons tend to be less 
isolated and have smaller $p_T$).   

We combine the signal and background templates taking into account the
above $W$ polarization dependent efficiencies. 
We use an unbinned likelihood method which
determines the $f_{0}$ and $f_{+}$ polarization fractions that best
correspond to the observed \ct distribution.   
A Gaussian constraint on the number of background events and a
Poisson constraint on the total number of observed events in data are 
included in the likelihood formula. We multiply the likelihoods 
for ``$b$-tag'' and ``non-tag'' subsamples to arrive at the final likelihood.
The method has been extensively tested in simulated samples across the
full range of physically possible values of $f_0$ and $f_+$ 
parameters. From these tests, we obtain small corrections to the
measured values of $f_0$ and $f_+$.

%
%
The determination of $W$-boson polarization fractions by our method is
sensitive to different sources of theoretical and experimental
uncertainties, such as  the MC simulated templates, the  jet
reconstruction algorithms, and jet corrections.  
 We have generated ensembles of simulated experiments in order to
 estimate these systematic uncertainties. 
One of the largest sources of systematic uncertainty comes from the
 jet energy scale (JES). We have studied this uncertainty by
changing the corrections by $\pm 1 \sigma$ of the JES
uncertainty~\cite{JES_NIM}.  
Another large systematic uncertainty is modeling of the signal
 which we estimate as variations in the ISR and FSR, using
different parton distribution functions (PDF) and different MC
generators (see~\cite{CDFtop_xs_dil} for details).
We estimate the systematic uncertainty due to the background template
shape by changing each individual background within 
its rate uncertainty thus changing the overall shape. We then combine
all these shifts (in quadrature) to obtain an overall background shape
uncertainty. The uncertainty in the total number of expected  
background events is taken into account in the fitting procedure where
the amount of background is allowed to float.
The method-specific systematic uncertainties are due to limited
statistics of the signal and background templates and are evaluated by
fluctuating the templates bin-by-bin.
An additional (small) uncertainty is due to the instantaneous
luminosity which determines the mean number of interactions per bunch
crossing.  
The systematic uncertainties are summarized in Table~\ref{tab:sys}, 
with the total systematic uncertainty of the measurement 
being the sum in quadrature of all the partial systematic
uncertainties from the various sources.

\begin{table}[htbp]
  \caption{Summary of systematic uncertainties.} \label{tab:sys}
  \begin{center}
    \begin{ruledtabular}
      \begin{tabular}{lcccc} 
        \multicolumn{1}{c}{Source }  &  $\Delta f_{0}^{1D}$ &  $\Delta f_{+}^{1D}$ & $\Delta f_{0}^{2D}$  & $\Delta f_{+}^{2D}$ \\ \hline 
        Jet energy scale    &  0.033  &  0.019  &  0.002  &  0.020 \\
        Generators          &  0.035  &  0.019  &  0.016  &  0.011 \\
        ISR/FSR             &  0.024  &  0.010  &  0.040  &  0.017 \\
        PDF                 &  0.010  &  0.003  &  0.025  &  0.009 \\
        Background shape    &  0.012  &  0.005  &  0.023  &  0.010 \\ 
        Template statistics &  & & &  \\
        \ \ \  Signal       &  0.010  &  0.005  &  0.024  &  0.012 \\
        \ \ \  Background   &  0.007  &  0.004  &  0.015  &  0.007 \\ 
        Instant. luminosity &  0.016  &  0.008  &  0.013  &  0.002 \\ \hline
        Total               &  0.059  &  0.031  &  0.063  &  0.034 \\
      \end{tabular}
    \end{ruledtabular}
  \end{center}
\end{table}
 
We assume a fixed top-quark mass of $m_{top} = 175$ \GeVc2 in our
dilepton measurement. However, as already noted, within the SM the
fraction of $W$-bosons with a given polarization directly depends on
the top-quark and $W$-boson masses. We do not include this effect 
in the systematic uncertainties for the result in the dilepton channel.
Rather, we provide the $m_{top}$-dependence of the reconstructed
fractions. 
We estimate that there is a linear shift in reconstructed $f_0$
 of $\pm(0.004 \pm 0.007)$  and $\pm (0.012 \pm 0.004)$  and in
 reconstructed $f_+$ of $\pm (0.005 \pm 0.004)$ and $\pm (0.006 \pm
 0.002)$  per  $\pm 1\ \GeVc2$ change in the top-quark mass for 2D and
 1D  measurements, respectively.    
In order to be close to the world average top-quark 
mass~\cite{cdfd0_topMassCombination}, we use the top-quark mass dependence
(presented above) to correct the result to a top-quark mass of 172.5 \GeVc2. 
 
%
%

\begin{figure}[htbp]
  \begin{center}
    \includegraphics[width=3.375in]{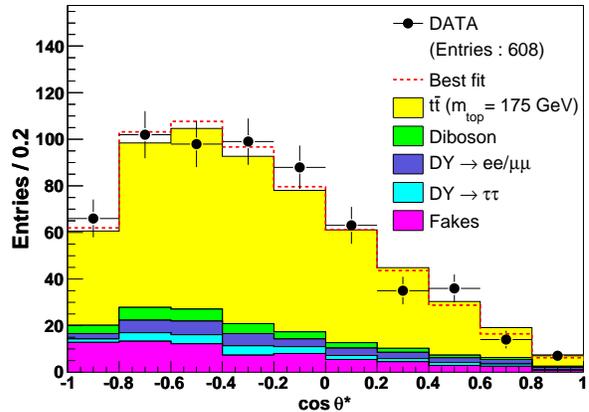}
    \caption{The \ct distributions for data, the expected SM $t\bar{t}$ signal 
      and background, and the best fit value.
      Each event has two entries in the histogram.}   
    \label{fig:cosTheta_data}
  \end{center} 
\end{figure}

There are 304 events (118 in ``$b$-tag'' and 186 in ``non-tag''
subsamples) passing dilepton selection and kinematic reconstruction,
consistent with the SM expectation of $284.3 \pm 22.7$ events.   
The comparison of \ct distribution between data and the expectations for SM
$t\bar{t}$ signal and background can be seen in
Fig.~\ref{fig:cosTheta_data}. There is a good agreement between data
and the SM expectation ($\chi^2 = 6.5$ for 9 degrees of freedom,
corresponding to a $p$-value of 69\%).  

We perform a model-independent simultaneous determination of
both $f_0$ and $f_+$ fractions: $f_0 = 0.70 ^{+0.18}_{-0.17} {\rm
  (stat)}$ and $f_+ = -0.09 \pm 0.09 {\rm (stat)}$. 
There is a strong negative correlation of $-0.88$ between the
statistical uncertainties of $f_0$ and $f_+$.  
We also measure each polarization fraction when the other is fixed to
its SM value. We measure  $f_0 = 0.56 \pm 0.09 {\rm (stat)}$ when
 $f_+$ is so fixed and measure $f_+ = -0.09 \pm 0.04 {\rm (stat)}$  when
 $f_0$ is so fixed. We also find $f_+  < 0.07$ at 95\% C.L. when $f_0$
 is so fixed following a Bayesian procedure assuming a constant a priori
 probability for $f_+$ across the physically possible range.

The CDF measurement performed in the single-lepton channel obtained the
following result~\cite{whelicityCDF}, assuming a top-quark mass of 172.5
\GeVc2: $f_0 = 0.90 \pm 0.11 {\rm (stat)}
\pm 0.06 {\rm (syst)}$ and $f_+= -0.19 \pm 0.07 {\rm (stat)} \pm 0.06
{\rm   (syst)}$ with the correlation of --0.59 between $f_0$ and $f_+$.
This is consistent with the result presented in this Letter. We
combine both results using the analytic best linear unbiased
estimator method~\cite{Lyons:1988,Valassi:2003}. We 
 include the systematic uncertainties corresponding to a $\pm 1.1$ \GeVc2 
uncertainty on $m_{top}$~\cite{cdfd0_topMassCombination}.   
The results of the dilepton and single-lepton channels are statistically
independent. There is a strong negative correlation of the statistical
uncertainty between the $f_0$ and $f_+$ observables for both channels
as mentioned above.  
 The systematic uncertainties are theoretically dominated and
are assumed to be 100\% correlated  between the measurements, with the
exception of the method-specific systematic uncertainties (signal and
background template statistics) which are treated as uncorrelated.
The luminosity-related systematic uncertainty applies only to the
dilepton measurement. For a given measurement, we assume that the $f_0$
and $f_+$ uncertainties are 100\% anti-correlated for each systematic
uncertainty category. Table~\ref{tab:combination}
presents the full correlation matrix between the measurements and their
weights in the combination. The combined result for the simultaneous
measurement is $f_0 =  0.84 \pm 0.09 {\rm (stat)} \pm 0.05  {\rm
  (syst)}$ and $f_{+}  =  -0.16 \pm 0.05  {\rm (stat)} \pm 0.04
{\rm (syst)}$. The combination has a $\chi^2$ value of 0.99 for two
degrees of freedom, corresponding to a p-value of 61\% for consistency
between the input measurements. The combined values of $f_0$ and $f_+$
have a correlation coefficient --0.81.
We also combine the measurements of one polarization
fraction when the other one is fixed to its SM expected value. In this
case, we arrive at $f_0 =  0.64 \pm 0.06 {\rm (stat)} \pm 0.05  {\rm
  (syst)}$ ($f_+$ is fixed) and $f_{+}  =  -0.07 \pm 0.02  {\rm
  (stat)} \pm 0.04 {\rm (syst)}$ ($f_0$ is fixed).
The combination for $f_0$ ($f_+$) has a $\chi^2$ of 1.04 (0.61) for one
degree of freedom, corresponding to a p-value of 31\% (44\%) for
consistency between the input measurements. 

\begin{table}
  \caption{ The correlation coefficients among the
    measurements and their weights in the $f_0$ and $f_+$ combined
    result. The results from single-lepton channel are labeled as
    `LJ', the dilepton results as `DIL'.}    
  \label{tab:combination}
  \begin{center}
    \begin{ruledtabular}
      \begin{tabular}{ldddddd} 
        Measurement  
        &  \multicolumn{4}{c}{ Correlation matrix}  
        &  \multicolumn{1}{c}{Weight } 
        &  \multicolumn{1}{c}{Weight } \\

        \multicolumn{1}{c}{ \ } &
        \multicolumn{1}{c}{\footnotesize LJ$f_0$ }    &
        \multicolumn{1}{c}{\footnotesize DIL$f_0$}    &
        \multicolumn{1}{c}{\footnotesize LJ$f_+$ }    &
        \multicolumn{1}{c}{\footnotesize DIL$f_+$}    &
        \multicolumn{1}{c}{\footnotesize for $f_0$ ($\%$)} & 
        \multicolumn{1}{c}{\footnotesize for $f_+$ ($\%$)} \\ \hline
        {\footnotesize LJ$f_0$  }   &    1  & & & &                             80.6  &  -21.8 \\
        {\footnotesize DIL$f_0$ }   &    0.13  &     1  &    & &                19.4  &   21.8 \\
        {\footnotesize LJ$f_+$  }   &   -0.72  &    -0.15  &    1  &  &         18.9  &   24.0 \\
        {\footnotesize DIL$f_+$ }   &   -0.12  &    -0.88  &    0.16  &    1 & -18.9  &   76.0 \\ 
      \end{tabular}
    \end{ruledtabular}
  \end{center}
\end{table}

%
%
To summarize, we have performed a measurement of $W$-boson polarization
fractions in top-quark dilepton decays. Our method is the first
model-independent measurement of the $W$ polarization in the
dilepton channel from CDF.
We have also combined our dilepton measurement with our previous
measurement in the single-lepton channel.
Our results are consistent with the SM expectations and
do not require the introduction of new physics. They agree with the
results obtained by the D0 and ATLAS collaborations~\cite{whelicityD0, whelicityATLAS} 
which are of comparable  precision.   

\begin{acknowledgments}
We thank the Fermilab staff and the technical staffs of the
participating institutions for their vital contributions. This work
was supported by the U.S. Department of Energy and National Science
Foundation; the Italian Istituto Nazionale di Fisica Nucleare; the
Ministry of Education, Culture, Sports, Science and Technology of
Japan; the Natural Sciences and Engineering Research Council of
Canada; the National Science Council of the Republic of China; the
Swiss National Science Foundation; the A.P. Sloan Foundation; the
Bundesministerium f\"ur Bildung und Forschung, Germany; the Korean
World Class University Program, the National Research Foundation of
Korea; the Science and Technology Facilities Council and the Royal
Society, UK; the Institut National de Physique Nucleaire et Physique
des Particules/CNRS; the Russian Foundation for Basic Research; the
Ministerio de Ciencia e Innovaci\'{o}n, and Programa
Consolider-Ingenio 2010, Spain; the Slovak R\&D Agency; the Academy of
Finland; and the Australian Research Council (ARC).  
\end{acknowledgments}


\begin{thebibliography}{99}
\bibitem{transverseEt} CDF uses a cylindrical coordinate system with
  the $z$ axis along the proton beam axis. Pseudorapidity is $\eta
  \equiv -{\rm ln} \tan (\theta/2)$, where $\theta$ is the polar angle relative
  to the proton beam direction and $\phi$  is the azimuthal angle
  while $p_T = |p| \sin \theta$, $E_T = E \sin \theta$.
\bibitem{muon}
  For the muons, transverse momentum $p_T$ rather than transverse
  energy $E_T$ is considered in the text.
\bibitem{missingTransverseEt} The missing transverse energy
  (${\not\!\!\vec{E}_T}$)   
  is defined by ${\not\!\!\vec{E}_T} = - \sum_{i} E_T^i \hat{n}_i$,
  where $i$ denotes calorimeter tower number with $|\eta| < 2.5$, and 
  $\hat{n}_i$ is a unit vector perpendicular to the beam
axis and pointing at the i$^{th}$ calorimeter tower. We also define
$\not\!\!E_T=|{\not\!\!\vec{E}_T}|$.
\bibitem{jetDefinition} A jet is defined as a fixed-cone cluster in the
  calorimeter with a cone size of 0.4 in
  azimuth-pseudorapidity ($\eta$) space.
\end{thebibliography}
\end{document}